\documentstyle[epsfig]{jaa}

\begin{document}
\title[TreePM: A code for Cosmological N-Body Simulations]{TreePM: A code for Cosmological N-Body Simulations}
\author[J.S.Bagla]{J.S.Bagla \\
Harish-Chandra Research Institute, Chhatnag Road, Jhunsi, \\
Allahabad 211019, INDIA \\
e-mail:jasjeet@mri.ernet.in}  

\volume{23}
\pubyear{2002}
\pagerange{\pageref{firstpage}--\pageref{lastpage}}
\date{Received 2002 June 13; accepted 2002 November 14}
\maketitle
\label{firstpage}

\begin{abstract}
We describe the TreePM method for carrying out large N-Body
simulations to study formation and evolution of the large scale
structure in the Universe.  This method is a combination of Barnes and
Hut tree code and Particle-Mesh code.  
It combines the automatic inclusion of periodic boundary conditions of
PM simulations with the high resolution of tree codes.  This is done
by splitting the gravitational force into a short range and a long
range component.  We describe the splitting of force between these two
parts.  We outline the key differences between TreePM and some other
N-Body methods. 
\end{abstract}


\section{Introduction}

Observations suggest that the present universe is populated by very
large structures like galaxies, clusters of galaxies etc.  Current
models for formation of these structures are based on the assumption
that gravitational amplification of density perturbations resulted in the
formation of large scale structures.  In absence of analytical
methods for computing quantities of interest, numerical simulations
are the only tool available for study of clustering in the non-linear
regime.  Last two decades have seen a rapid development of
techniques and computing power for cosmological simulations and the
results of these simulations have provided valuable insight into the
study of structure formation.  

The simplest N-Body method that has been used for studying clustering of
large scale structure is the Particle Mesh method (PM hereafter).  The
genesis of this method is in the realisation that the Poisson equation
is an algebraic equation in Fourier space, hence if we have a tool for
switching to Fourier space and back, we can calculate the
gravitational potential and the force with very little effort.  It
has two elegant features in that it provides periodic boundary
conditions by default, and the force is softened naturally so as to
ensure collisionless evolution of the particle distribution.  However,
softening of force done at grid scale implies that the force
resolution is very poor.  This limits the dynamic range
over which we can trust the results of the code between a few grid
cells and about a quarter of the simulation box~(Bouchet and Kandrup,
1985; Bagla and Padmanabhan, 1997.  Many efforts have been made to get around
this problem, mainly in the form of P$^3$M (Particle-Particle Particle
Mesh) codes~(Efstathiou et al, 1985; Couchman 1991).  In these codes,
the force computed by 
the particle mesh part of the code is supplemented by adding the short
range contribution of nearby particles, to improve force resolution.
The main problem with this approach is that the particle-particle
summation of the short range force takes a lot of time in highly
clustered situations.  Another, more subtle problem is 
that the force computed using the PM method has anisotropies and
errors in force at grid scale -- these errors are still present in the force
calculated by combining the PM force with short range
corrections~(Bouchet and Kandrup, 1985). 

A completely different approach to the problem of computing force are
codes based on the tree method.  In this approach we consider groups of
particles at a large distance to be a single entity and compute the
force due to the group rather than sum over individual particles.  There
are different ways of defining a group, but by far the most popular
method is that due to Barnes and Hut~(1986).  Applications of this
method to Cosmological simulations require including periodic boundary
conditions.  This has been done using Ewald's method~(Ewald, 1921;
Rybicki, 1986; Hernquist, Bouchet and Suto, 1991; Springel, Yoshida
and White, 2001).  Ewald's 
method is used to tabulate the correction to the force due to periodic
boundary conditions.  This correction term is stored on a grid (in
relative separation of a pair of particles) and the interpolated value
is added to the pairwise force.  

Some attempts have been made to combine the high resolution of a tree
code with the natural inclusion of periodic boundary conditions in a PM
code by simply extending the P$^3$M method and replacing the
particle-particle part for short range correction with a local
tree~(Xu, 1995).  

In this paper we present a hybrid N-Body method that attempts to
combine the good features of the PM and the tree method, while
avoiding the problems of the P$^3$M and the TPM methods.  Our approach
is to divide force into long and short range
components using partitioning of unity, instead of taking the PM force
as given.  This allows us greater control over errors, as we shall
see below. 

The plan of the paper is as follows:  \S{2} introduces the basic
formalism of both the tree and PM codes.  \S{2.3} gives the
mathematical model for the TreePM code.  We analyse errors in force
for the TreePM code in \S{3}.  Computational requirements of our
implementation of the TreePM code are discussed in \S{4}.  A
discussion of the relative merits of the TreePM method with respect to
other N-Body methods follows in \S{5}. 

\section{The TreePM Method}

\subsection{Tree Code}

We use the approach followed by Barnes and Hut~(1986).  In this, 
the simulation volume is taken to be a cube.  The tree structure is
built out of cells and particles.  Cells may 
contain smaller cells (subcells) within them.  Subcells can have even
smaller cells within them, or they can contain a particle.  We start
with the simulation volume and add particles to it.  If two particles
end up in the same subcell, the subcell is geometrically divided into
smaller subcells until each subcell contains either subcells or at
most one particle.  The cubic simulation volume is the root cell.  In
three dimensions, each cubic cell is divided into eight cubic
subcells.  Cells, as structures, have attributes like total mass,
location of centre of mass and pointers to subcells.  Particles, on
the other hand have the traditional attributes like position, velocity
and mass. More details can be found in the original paper~(Barnes and
Hut, 1986).

Force on a particle is computed by adding contribution of other
particles or of cells.  A cell that is sufficiently far away can be
considered as a single entity and we can just add the force due to the
total mass contained in the cell from its centre of mass.  If the cell
is not sufficiently far away then we must consider its constituents,
subcells and particles.  Whether a cell can be accepted as a single
entity for force calculation is decided by the cell acceptance
criterion (CAC).  We compute the ratio of the size of the cell $d$
and the distance $r$ from the particle in question to its centre of
mass and compare it with a threshold value
\begin{equation}
\theta = \frac{d}{r} \leq \theta_c  \label{trwalk}
\end{equation}
The error in force increases with $\theta_c$.  There
are some potentially serious problems associated with using $\theta_c
\geq 1/\sqrt{3}$, a discussion of these is given in Salmon and
Warren~(1994).  One can also work with completely different
definitions of the CAC~(Salmon and Warren, 1994; Springel, Yoshida and
White, 2001).  Irrespective of the 
criterion used, the number of terms that contribute to the force on a
particle is much smaller than the total number of particles, and this
is where a tree code gains in terms of speed over direct summation. 

We will use the Barnes and Hut tree code and we include periodic boundary
conditions for computing the short range force of particles near the
boundaries of the simulation cube.  Another change to the standard
tree walk is that we do not consider cells that do not have any
spatial overlap with the region within which the short range force is
calculated.  We also use an optimisation technique to speed up force
calculation~(Barnes, 1990). 

\subsection{Particle Mesh Code}

A PM code is the obvious choice for computing long range interactions.
Much has been written about the use of these in cosmological
simulations (e.g., see Hockney and Eastwood, 1988) so we will not go
into details here.  PM codes solve for the gravitational potential
in the Fourier space.  These use Fast Fourier Transforms (FFT) to
compute Fourier transforms, and as FFT requires data to be defined on
a regular grid the concept of mesh is introduced.  The density field
represented by particles is interpolated onto the mesh.  Poisson
equation is solved in Fourier space and an inverse transform gives the
potential (or force) on the grid.  This is then differentiated and
interpolated to the position 
of each particle in order to calculate the displacements.  Use of a
grid implies that forces are not accurate at the scale smaller than
the grid cells.  A discussion of errors in force in a PM code can be
found in Efstathiou et al~(1985) and elsewhere~(Bouchet and Kandrup,
1985; Bagla and Padmanabhan, 1997).  The error in force can be very
large at small scales but it drops to an acceptable number beyond a
few grid cells, and is negligible at large scales.

We use the Cloud-in-Cell weight function for interpolation.  We solve
the Poisson equation using the natural kernel, $-1/k^2$; this is called
the poor man's Poisson solver~(Hockney and Eastwood, 1988).  We
compute the gradient of the potential in Fourier space.

\subsection{TreePM Code}

We now turn to the question of combining the tree and the PM code.  We
wish to split the inverse square force into a long range force and a
short range force.  The gravitational potential can be split into two
parts in Fourier space~(Ewald, 1921).  
\begin{eqnarray}
\varphi_k &=& - \frac{4 \pi G \varrho_k}{k^2} \label{pm_std}\\
 &=& - \frac{4 \pi G \varrho_k}{k^2} \exp\left(-k^2 r_s^2\right)  -
 \frac{4 \pi G \varrho_k}{k^2} \left(1 - \exp\left(-k^2
 r_s^2\right)\right)\nonumber\\ 
 &=& \varphi_k^l + \varphi_k^s \nonumber \\ 
\varphi_k^l &=& - \frac{4 \pi G \varrho_k}{k^2} \exp\left(-k^2
 r_s^2\right) \label{longr}\\
\varphi_k^s &=& - \frac{4 \pi G \varrho_k}{k^2} \left(1 - \exp\left(-k^2
 r_s^2\right)\right) \label{shortr}
\end{eqnarray}
where $\varphi^l$ and $\varphi^s$ are the long range and the short range
potentials, respectively.  The splitting is done at the scale $r_s$.  $G$
is the gravitational coupling constant and $\varrho$ is density.  The
expression for the short range force in real space is:
\begin{equation}
{\bf f}^s({\bf r}) = - \frac{G m {\bf r}}{r^3} \left({\rm
erfc}\left(\frac{r}{2 r_s}\right) + \frac{r}{r_s \sqrt{\pi}}
\exp\left(-\frac{r^2}{4 r_s^2}\right)\right) \label{fshort}
\end{equation}
Here, ${\rm erfc}$ is the complementary error function.
These equations describe the mathematical model for force in the
TreePM code.  The long range potential is computed in the Fourier
space, just as in a PM code, but using eqn.(\ref{longr}) instead of
eqn.(\ref{pm_std}).  This potential is then used to compute the long
range force.  The short range force is computed directly in real space
using eqn.(\ref{fshort}).  In the TreePM method this is computed using
the tree approximation.  The short range force falls rapidly at scales 
$r \gg r_s$, and hence we need to take this into account only in a
small region around each particle. 

\begin{figure}
\epsfxsize=4truein\epsfbox[39 27 513 506]{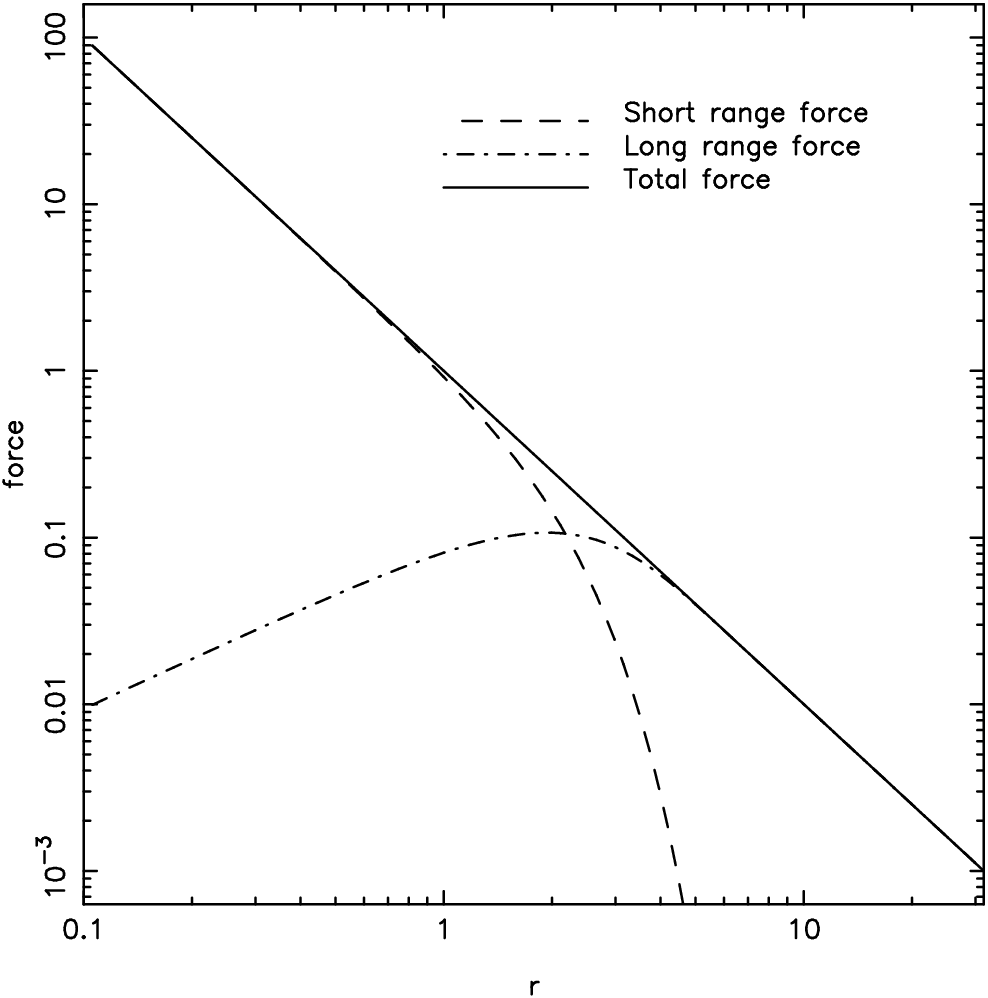}
\caption{This figure shows the long and the short range force as a
function of scale.  The inverse square force is shown by the thick
line, the long range force by dot-dashed line and the short range
force by the dashed line. We have taken $r_s=1$ here.}
\end{figure}

We have plotted the long range and the short range
force (eqn.(\ref{fshort})) as a function of $r$ in fig.1 to show
their dependence on scale.  We have chosen $r_s=1$ here.  The short
range force closely 
follows the total force up to about $2 r_s$ and then falls rapidly,
its magnitude falls below $1\%$ of the total force by $5 r_s$.  The long
range force reaches a peak around $2 r_s$.  It makes up most of the
total force beyond $3.5 r_s$.  It falls with scale below $2 r_s$,
becoming negligible below $r_s / 2$.

Evaluation of special functions for calculating the short range
force can be time consuming.  To save time, we compute an array 
containing the magnitude of the short range force.  The force between
any two objects, particle-cell or particle-particle, is computed by
linearly interpolating between the nearby array elements multiplied 
by the unit vector ${\bf r}$.  It is necessary for the array to
sample the force at sufficiently closely spaced values of $r$ in order
to keep error in interpolation small.

\section{Error Estimation}

In this section we will study errors in force introduced by various
components of the TreePM code.  We will only list salient points here
and the reader is referred to a more comprehensive study for
details~(Bagla and Ray, 2002). 

We start by estimating the error in force due to one particle.  The
long range force of a particle is calculated using the PM method, but
using eqn.(\ref{longr}) instead of eqn.(\ref{pm_std}).  The cutoff at
high wave numbers largely removes the effect of the grid and we find
that the dispersion in the long range force is very small, e.g. for
$r_s \geq 1$ grid length the dispersion is smaller than $1\%$ of the
total force at all scales.  There is a systematic offset in the long
range force that is larger than the dispersion.  This offset is
induced by the interpolating function, and can be corrected~(White,
2000; Bagla and Ray, 2002) by de-convolving the square of the
interpolating function (we need to interpolate twice).  This
deconvolution does not affect the dispersion in any significant
manner.

There are no errors in computing the short range force for one
particle, hence the only source of errors is in the calculation of
the long range force in this case.  All the errors arise due to
anisotropies in the long range force.  The errors in the long range
force increase as we approach small scales, but the contribution of
the long range force to the total force falls sharply below $2r_s$ and
hence the errors also drop rapidly.  There is a peak in errors
around $2r_s$--$3r_s$, and for $r_s=1$ maximum rms error in force of
one particle is $1\%$ of the total force.

In calculating the total force, we added the short range force to the
long range force at all scales.  However, this is not necessary as
beyond some scale, the contribution of small scale force to the total
force drops to a negligible fraction of the total force.  We will call
the scale upto which we add the small scale force as $r_{cut}$.  The
short range force is just below $1\%$ of
the total force at $r_{cut}=5r_s$.  We choose this value of $r_{cut}$
for the TreePM code.

\begin{figure}
\epsfxsize=4truein\epsfbox[38 27 513 506]{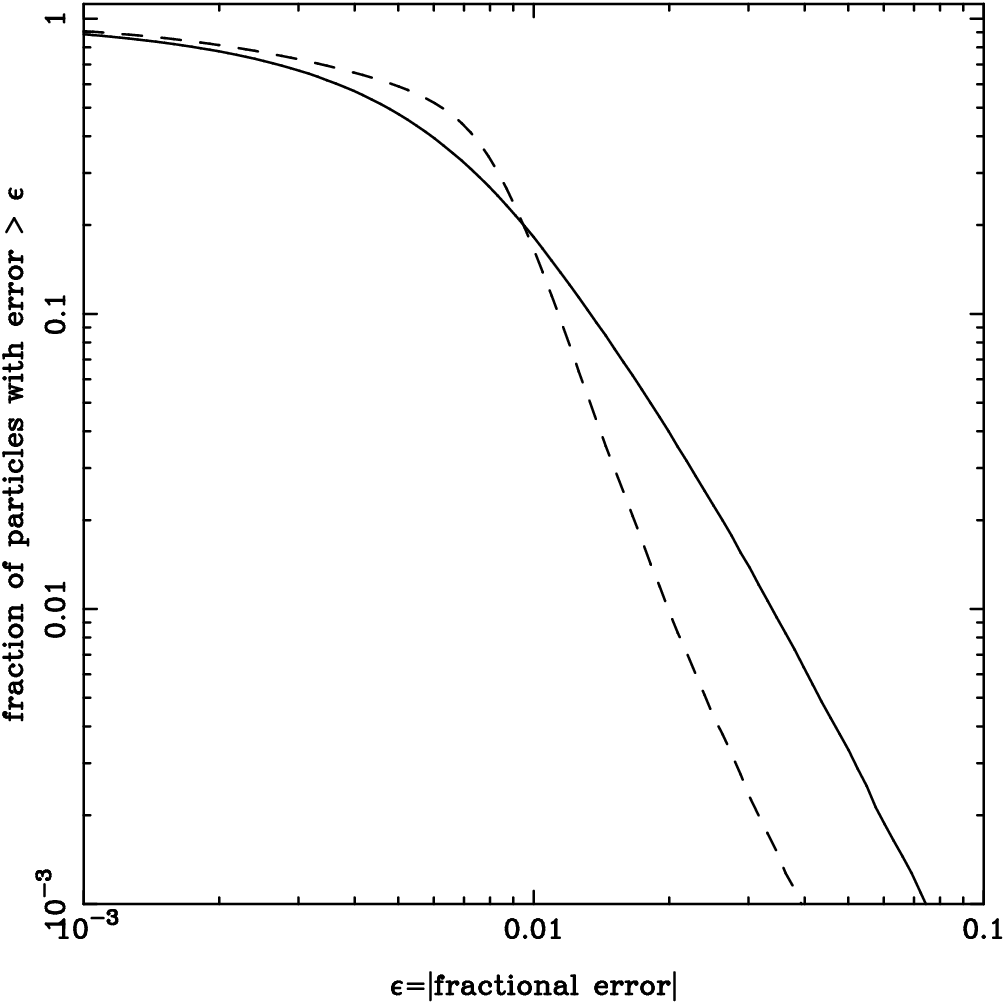}
\caption{This figure shows the distribution of errors.  The variation
of the fraction of particles with error greater than a threshold, as a
function of the threshold error is plotted.  Thick line marks
the error for a homogeneous distribution of particles and the dashed
line shows the same for a clumpy distribution.  These errors were
measured with respect to a reference force, determined with a very
conservative value of $r_s$, $r_{cut}$ and $\theta_c$.  This panel
shows that $99\%$ of the particles have fractional error in force that
is less than $3.5\%$ for the homogeneous distribution and around
$2\%$ for the clumpy distribution.}
\end{figure}

The other source of error is the tree approximation that we use for
computing the short range force.  The first correction term is due to
the quadrapole moment of the particle distribution in the cell,
however the magnitude of this error is larger than in the inverse
square force due to a more rapid variation in force with distance.  In
the worst case, this error can be more than twice the error in the
corresponding case of inverse square force~(Bagla and Ray, 2002).  In
more generic cases, errors due to this effect tend to cancel out and
the net error is small. 

Apart from this effect, there is also a dispersion
introduced by the tree approximation.  The magnitude of this
dispersion varies monotonically with $\theta_c$.

One factor that we have to weigh in is that the execution time is
small for large $\theta_c$ and small $r_{cut}$.  Given these
considerations, the obvious solution is to choose the smallest $r_s$
and the largest $\theta_c$ that gives us a sufficiently accurate force
field. 

It is important to estimate the errors in a realistic situation, even
though we do not expect errors to add up coherently in most
situations.  We test errors for two distributions of particles: a
homogeneous distribution and a clumpy distribution.  For the
homogeneous distribution, we use randomly distributed particles in a
box.  We use $262144$ particles in a $64^3$ box for this distribution.
We compute the force using a reference setup ($r_s=4$, $r_{cut}=6
r_s$, $\theta_c=0$) and the setup we wish to test ($r_s=1$, $r_{cut}=5
r_s$, $\theta_c=0.5$).  It can be shown that the errors in the
reference setup are well below $0.5\%$ for the entire range of
scales~(Bagla and Ray, 2002).  We compute the fractional error in
force acting on each particle, this is defined as, 
\begin{equation}
\epsilon = \frac{\left\vert {\bf f} - {\bf f}_{ref}
\right\vert}{\left\vert {\bf f}_{ref} \right\vert}  .
\end{equation}
Fig.2 shows the cumulative distribution of fractional errors.  The
curves show the fraction of particles with error greater than 
$\epsilon$.  The thick line shows this for the homogeneous
distribution.  Error $\epsilon$ for $99\%$ of particles is less than
$3.5\%$.  Results for the clumpy distribution of particles are shown
by the dashed line.  We used the output of a CDM simulation (fig.3a)
run with the TreePM code.  Errors in this case are much smaller, as
compared to the homogeneous distribution, as in the case of tree
code~(Hernquist, Bouchet and Suto, 1991).  Error $\epsilon$ for $99\%$
of particles is around $2\%$, as compared to $3.5\%$ for the
homogeneous distribution.  

\begin{figure}
\epsfxsize=3.2truein\epsfbox[35 25 512 505]{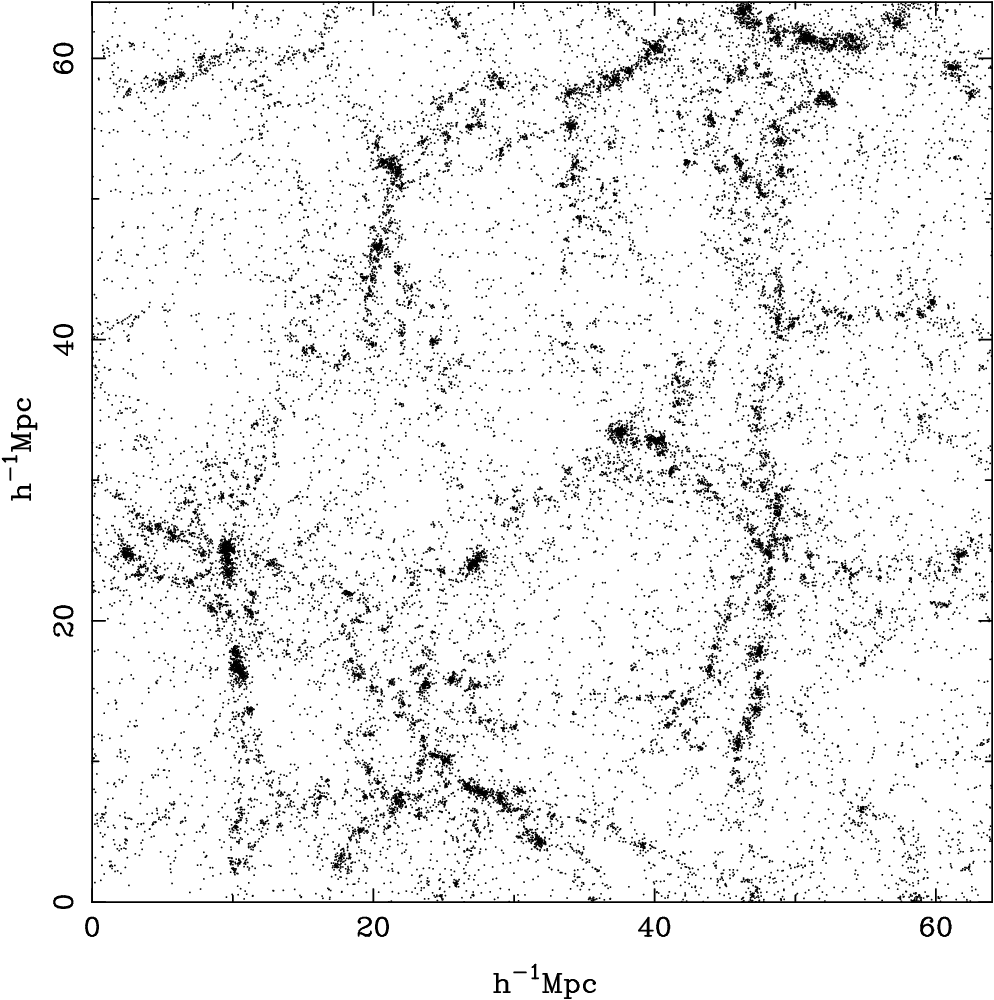}
\epsfxsize=3.2truein\epsfbox[35 25 512 535]{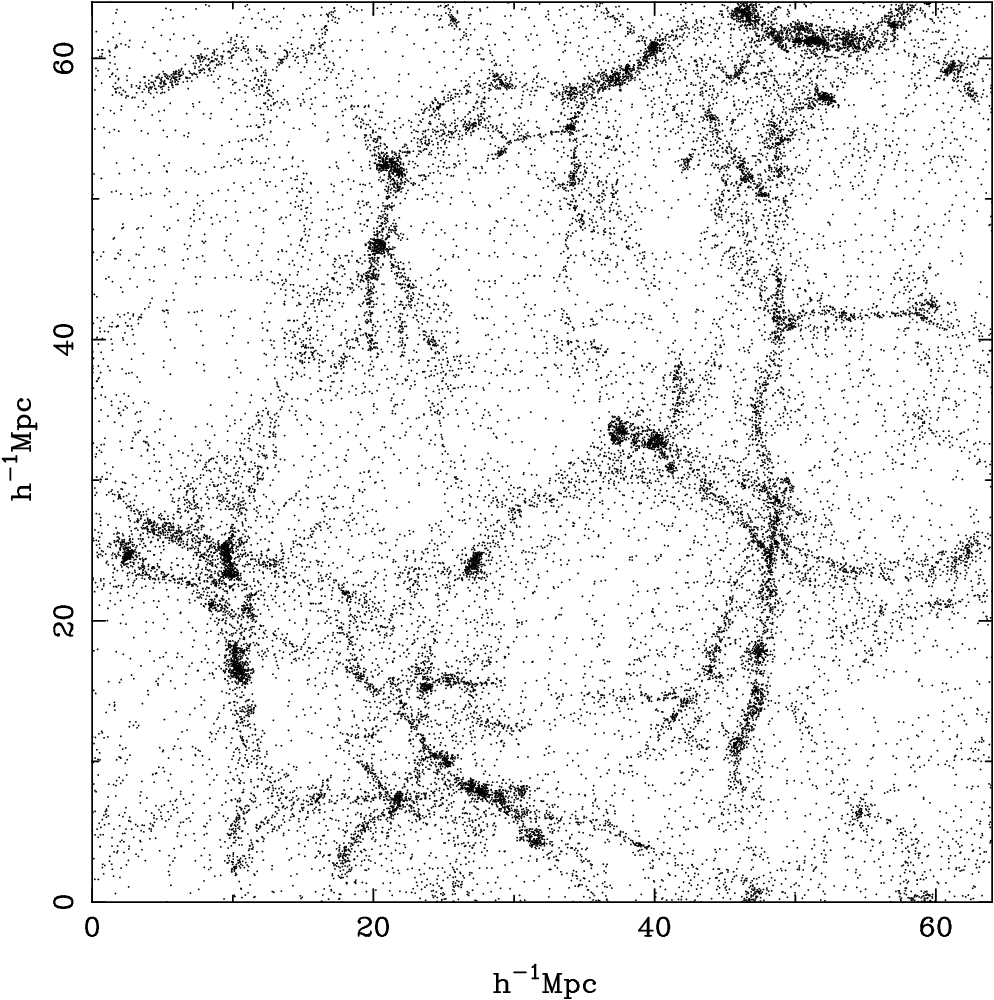}
\caption{This figure shows a slice from a simulation of the sCDM
model.  The top panel shows the slice from the TreePM simulation.  For
comparison, we have included the same slice from a PM simulation of
the same initial conditions in the lower panel.  The large scale
structures are the same in 
the two but there are significant differences at small scales.  The
halos are much more compact in the TreePM simulation, and large halos
show more substructure.  This is to be expected because of the
superior resolution of the TreePM code.}
\end{figure}

There are two noteworthy features of this figure.  One is that the
error for the homogeneous distribution is higher.  The main reason for
this is similar to that in tree codes, though the effect is much
smaller here.  When we are dealing with a homogeneous distribution,
the total force on each particle is very small because forces due to
nearly identical mass distributions on opposite sides cancel out.
This near cancellation of large numbers gives rise to errors that
decrease as the net result of these cancellations grows.  In a tree
code, we calculate the force due to all the particles in the
simulation box whereas in the TreePM method we add up the contribution
of only those within a sphere of radius $r_{cut}$.  This is the
reason for the difference in these two curves being much
less pronounced than the corresponding curves for the tree
code~(Hernquist, Bouchet and Suto, 1991). 

The other feature is that the shape of the curves for the homogeneous
distribution and the clumpy distribution is different.  This is
because we begin to see the effect of the error due to tree
approximation in case of clumpy distribution.  In case of the
homogeneous distribution, the distribution of particles is close to
isotropic around any given particle and hence the error cancels out.
This error can be controlled by reducing $\theta_c$.   

We end this section with a brief comparison of the TreePM code with a PM
code.  We ran a simulation of the sCDM model ($262144$ particles,
$64$h$^{-1}$Mpc box) with a PM code (Bagla and Padmanabhan, 1997) and
with the TreePM 
code discussed here.  Fig.3 shows a slice from these simulations; fig.3a
shows the simulation with the TreePM code and fig.3b shows the same for
a PM code.  The large scale structures are the same in
the two but there are significant differences at small scales.  The
halos are much more compact in the TreePM simulation, and large halos
show more substructure.  These differences are also clear in the two
point correlation function $\bar\xi(r)$ plotted in fig.4.  The thick
line shows the correlation from the TreePM simulation and the dashed
line shows the same for the PM simulation.  As expected from fig.3 and
from general considerations, the correlation function in the TreePM
simulation matches with that from the PM simulation at large scales, but
at small scales, the TreePM simulation has a higher correlation
function.

\begin{figure}
\epsfxsize=4truein\epsfbox[38 25 512 505]{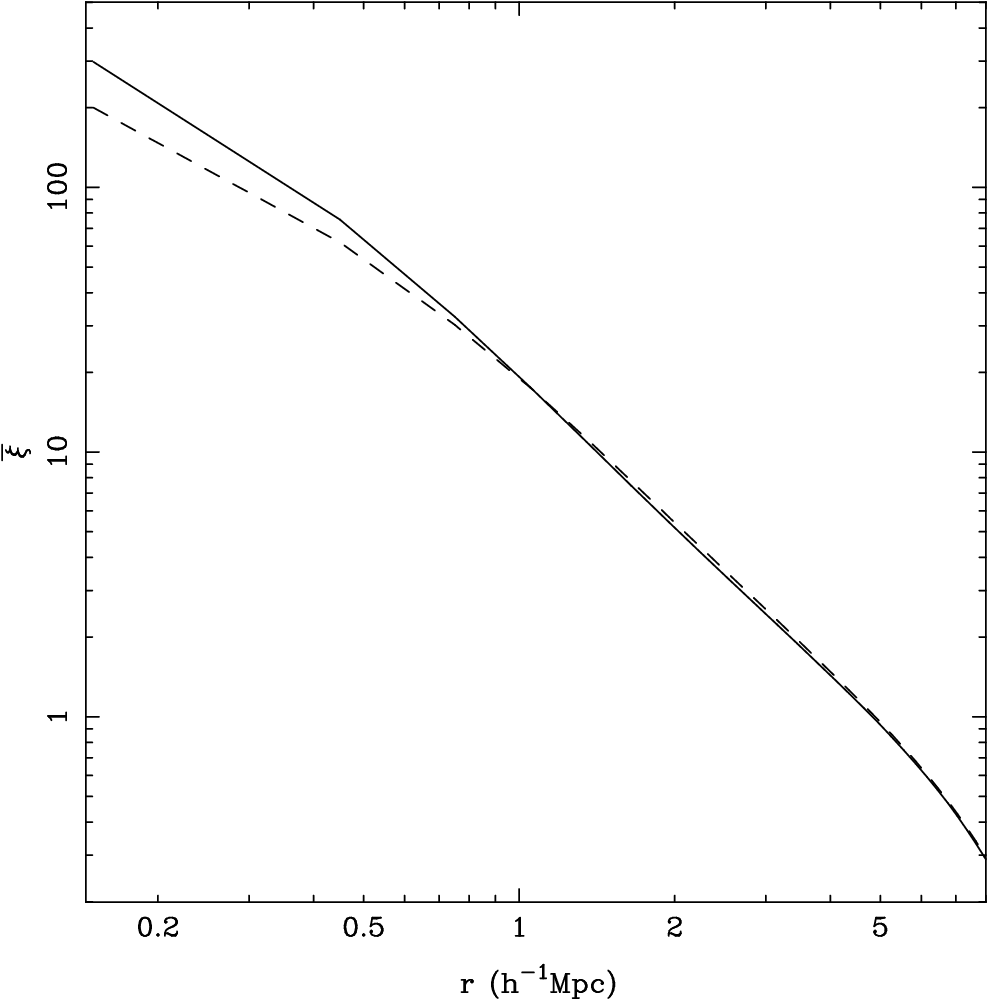}
\caption{This figure shows the averaged correlation function
$\bar\xi(r)$ as a function of scale.  The thick line shows this
quantity for the TreePM simulations and the dashed line shows the same
for the PM simulation.  These two match at large scales
but the PM simulation underestimates the clustering at small scales.}
\end{figure}

We have checked the accuracy of evolution by checking the rate of
growth for the correlation function in the linear regime and also by
looking for scale invariance of the correlation function for power law
models.  For more details please see~(Bagla and Ray, 2002).

\section{Computational Resources}

In this section, we describe the computational resources required for
the present implementation of the TreePM code.  Given that we have
combined the tree and the PM code, the memory requirement is obviously
greater than that for either one code.  We need four arrays for the PM
part, the potential and the force.  The rest is exactly the same as a
standard Barnes and Hut tree code.  With efficient memory management, we
need less than $160$MB of RAM for a simulation with $128^3$ particles in
a $128^3$ mesh for most part.  In absence of memory management, this
requirement can go up to 250MB.  These are the numbers for floating
point numbers, if we use double precision variables then this
requirement goes up by a factor of two.

\begin{table}
\begin{tabular}{||l|l|l|l|l|l||}
\hline
$N_{particle}$ & time & time & time & time & time \\
 & (ms)& (ms) & (ms) & (ms) & \\
\hline
 & TreePM & TreePM & TreePM & TreePM & tree \\
\hline
 & unclustered & unclustered & unclustered & clustered & unclustered \\
\hline
 & P-4 & PIII & Alpha & Alpha & Alpha \\
\hline
$32768$ & & & 0.57 & 0.59 & 2.94 \\
\hline
$262144$ & & & 0.78 & 0.80 & 3.75 \\
\hline
$2097152$ & 0.34 & 0.89 & 1.22 & 1.28 & 6.03 \\
\hline
\end{tabular}
\vspace{15pt}
\caption{Time taken by the code, per time step per particle.  Column~1
lists the number of particles.  Column~2, 3, 4 and 5 list the time taken
(per time step per particle) by the TreePM code for an unclustered and
a clustered particle distribution.  Column~6 lists the same number for
a tree code for an unclustered distribution of particles.  All the
times are in milli~seconds.}
\end{table}

Table 1 lists the time required per time step per particle for three
values of the number of particles.  These were run on a 533MHz
Alpha workstation (EV5) and compiled with the native F90 compiler, a
$1$GHz Pentium III desktop or a $1.6$GHz P-4 and compiled with the Intel
F90 compiler. 
Column~1 lists the number of particles and col.2, 3 and 4 list the
time per step per particle for an unclustered distribution.  This
number increases much slower than the total number of particles, as
expected from the theoretical scaling of $O(N\ln{N})$.

Column 5 of table gives the same number for a highly clustered
particle distribution, similar in clustering strength to that shown in
fig.3.  Column~6 lists the time per step per particle taken by
the tree code for the particle distribution used in col.4.  It is
clear that the TreePM code is faster than the tree code by a factor of
about $4.5$.  It is also clear that this code performs well even on
inexpensive hardware.

The performance of this code can be improved further by including
features like individual time steps for particles.  It is expected
that adding individual time steps will improve the performance by a
factor of two or more.

\section{Comparison with other Methods}

Amongst other codes that try to augment the performance of PM codes
are the P$^3$M~(Efstathiou et al, 1985; Couchman, 1991) codes and the
TPM code~(Xu, 1995).  Following subsections compare TreePM with these
codes. 

\subsection{P$^3$M and AP$^3$M}

There are two main differences between P$^3$M codes~(Efstathiou et al,
1985; Couchman, 1991) and 
the TreePM code presented here.  One is that most P$^3$M codes use the
natural cutoff provided by the grid for the long range force,
i.e. these take the PM force to be the long range force.  Hence errors
in the PM force are present in the P$^3$M force.  In contrast, the
TreePM code uses an explicit cutoff that allows us to limit errors
near the grid scale. 

The second difference is in terms of the time taken for the adding the
short range correction as a function of clustering.  In both
instances, the short range force is added for particles within a fixed
radius $r_{cut}$.  This process is of order $O(N n r_{cut}^3 (1 +
\bar\xi(r_{cut})) )$ for the P$^3$M method, where $N$ is the number of
particles in the simulation, $n$ is the number density of particles
and $\bar\xi(r_{cut})$ is the average number
of excess particles around a particle, here excess is measured
compared to a homogeneous distribution of particles with the same
number density.  At early times this reduces to $O(N n r_{cut}^3)$,
but at late times, when the density field has become highly non-linear
($\bar\xi(r_{cut}) \gg 1$), it becomes $O(N n r_{cut}^3
\bar\xi(r_{cut}))$.  As the density field becomes more and more
clumpy, the number of operations required for computing the short
range force increase rapidly.  This is to be compared with the number
of operations required for adding the short range correction in
the TreePM code: $O(N \log(n r_{cut}^3 (1 + \bar\xi(r_{cut}))) )$.
The linear and the non-linear limits of this expression are $O(N
\log(n r_{cut}^3))$ and $O(N \log(n r_{cut}^3 \bar\xi(r_{cut})))$,
respectively.  Thus the variation in the number of operations with
increase in clustering is much less for TreePM code than a P$^3$M code.
The problem is not as severe as outlined for the Adaptive P$^3$M
code~(Couchman, 1991) but it still persists.  Therefore the TreePM code has a
clear advantage over the P$^3$M and AP$^3$M code for simulations of
models where $\bar\xi(r_{cut})$ is very large. 

In turn, P$^3$M codes have one significant advantage over TreePM,
these require much less memory.  This gives P$^3$M codes an advantage
on small machines and for simulations of models where
$\bar\xi(r_{cut})$ is not much larger than unity. 

\subsection{TPM}

Before we go into the differences between the TreePM and TPM methods,
we would like to summarise the TPM method~(Xu, 1995) here.  

The TPM method is an extension of the P$^3$M method in that the PM
force is taken to be the long range force and a short range force is
added to it.  Tree method is used for adding the short range
correction instead of the particle-particle method.  There are some
further differences, e.g. correction is added only for particles in
high density regions implying that the resolution is non-uniform.  At
each time step, high density regions are identified and a local tree
is constructed in each of these regions for computing the short range
correction.  Thus, there are two clear differences between the TreePM
and the TPM method:
\begin{itemize}
\item The TPM code uses the usual PM force to describe the long range
component.  In contrast, the TreePM code uses an explicit cutoff
($r_s$).
\item TreePM treats all the particles on an equal footing, we
compute the short range (eqn(\ref{fshort})) and the long range force
for each particle.  In the TPM code, the short
range force is computed only for particles in the high density
regions.
\end{itemize}

\section{Discussion}

Preceeding sections show that we have developed a new method for doing
cosmological N-Body simulations with a clean mathematical model.  The
model splits force into long and short range forces using a parameter
$r_s$.  By choosing this parameter judiciously, in conjunction with
two other parameters that arise in the implementation of this model
($r_{cut}$ and $\theta_c$) we can obtain a configuration that matches
our requirements for the error budget. 

It is possible to devise a more complex scheme for splitting the force
into two parts but the one we have chosen seems to be the optimal
scheme from the point of view of errors in force calculation as well
as CPU time~(Bagla and Ray, 2002). 

Apart from improving control over errors, the TreePM code also leads
to a significant gain in speed over the traditional tree code.  

TreePM code is also amenable to parallelisation along the lines
of~(Dubinski, 1996), and is likely to scale well because the
communication overhead is much more limited.  Work in this direction is
in progress and will be reported elsewhere~(Bagla, 2002).

\section*{Acknowledgement}

I would like to thank Rupert Croft, Lars Hernquist, Suryadeep Ray,
Volker Springel and Martin White for insightful comments and
discussions.  Part of the work reported in this paper was done while
the author was at the Harvard-Smithsonian Center for Astrophysics.

\label{lastpage}

\end{document}